\documentclass[proof]{WileyASNA-v1}

\articletype{Conference Paper}%

\received{xxxxxxxxxx}
\revised{xxxxxxxxxx}
\accepted{xxxxxxxxxx}

\def \bea {\begin{eqnarray}}
\def \eea {\end{eqnarray}}
\def \nn   {\nonumber}
\begin{document}

\title{A minimal Length Uncertainty Approach to Cosmological Constant Problem}

\author[1]{Abdel Magied DIAB*}
\address[1]{\orgdiv{Modern University for Technology and Information (MTI)}, \orgname{Faculty of Engineering}, \orgaddress{\state{11571 Cairo}, \country{Egypt.}}}

\author[2,3]{Abdel Nasser TAWFIK$\dagger$}

\address[2]{\orgdiv{Nile University}, \orgname{Egyptian Center for Theoretical Physics (ECTP)}, \orgaddress{\state{12588 Giza}, \country{Egypt.}}}

\address[3]{\orgdiv{Goethe University}, \orgname{Institute for Theoretical Physics}, \orgaddress{\state{D-60438, Frankfurt am Main}, \country{Germany.}}}

\corres{*\email{a.diab@eng.mti.edu.eg} \\  $\dagger$\email{tawfik@itp.uni-frankfurt.de}}

\abstract{Based on quantum mechanical framework for the minimal length uncertainty, we demonstrate that the generalized uncertainty principle (GUP) parameter could be best constrained by recent gravitational waves observations on one hand. On other hand this suggests modified dispersion relations (MDRs) enabling an estimation for the difference between the group velocity of gravitons and that of photons. Utilizing features of the UV/IR correspondence and the obvious similarities between GUP (including non-gravitating and gravitating impacts on Heisenberg uncertainty principle) and the discrepancy between the theoretical and the observed cosmological constant (apparently manifesting gravitational influences on the vacuum energy density), we suggest a possible solution for the cosmological constant problem.}

%\pacs{04.30.-w, 04.60.-m, 02.40.Gh, 98.80.Es}

\keywords{Gravitational waves, Quantum gravity, Noncommutative geometry, Observational cosmology}

%\jnlcitation{\cname{%
%\author{Williams K.}, 
%\author{B. Hoskins}, 
%\author{R. Lee}, 
%\author{G. Masato}, and 
%\author{T. Woollings}} (\cyear{2016}),  
%\cjournal{Q.J.R. Meteorol. Soc.}, 
%\cvol{2017;00:1--6}.}

\fundingInfo{EXTP-2020-14 and WLCAPP-2020-14}

\maketitle

\footnotetext{\textbf{Abbreviations:} GR, general relativity; EFE, Einstein field equations; GUP, generalized uncertainty principle; HUP, Heisenberg uncertainty principle; UV/IR correspondence, Ultraviolet/Infrared correspondence; LIGO, Laser Interferometer Gravitational-Wave Observatory; GW, gravitational waves}

\section{Introduction}
\label{intro}

In his theory of general relativity (GR), Einstein proposed a cosmological constant $\Lambda$ to be included in the field equation (EFE) \cite{Einstein1917As} in order to secure a static evolution of the universe \cite{Tawfik:2011mw, Tawfik:2008cd}. With the discovery of \cite{Hubble:1929ig} that the universe is likely expanding, such models have been discarded. The Hubble's observation was also confirmed by the Friedmann solutions of EFE in absence of $\Lambda$-term \cite{Friedman:1922kd}. Consequently, $\Lambda$ was also removed from EFE. Nevertheless later on, $\Lambda$ came back several times even by Einstein himself as it was realised that all models of static universe become unstable to small perturbations \cite{Mulryne:2005ef, Wu:2009ah}, especially that the universe apparently has a static origin, the the initial singularity of the expanding universe. In ealry nineteth of the last century, the observations of type-Ia high redshift supernovae \cite{Riess:1998cb, Perlmutter:1998np} suggested that the expanding universe seems even accelerating as well with a small value of the $\Lambda$-term, which strongly referred to negative pressure \cite{Garriga:1999bf, Martel:1997vi}. This is not the only puzzle with $\Lambda$.

The disagreement between estimated values of $\Lambda$ in theory and observation represents another mystery in physics. In $2018$, PLANCK collaboration \cite{Aghanim:2018eyx} provided us with so-far the most precise observational value $\Lambda_{obs}=10^{-47}$GeV$^4/(\hbar c)^3$ \cite{Aghanim:2018eyx}. On the other hand, the theoretical estimation suggests $\Lambda_{the}=10^{+74}$GeV$^4/(\hbar c)^3$ \cite{Zeldovich:1968ehl, Weinberg:1988cp}. There is a difference of $121$ orders of magnitude. The goal of present work is suggesting an interpretation based on affordable consistency between theoretical and observational $\Lambda$. We utilize some features of the UV/IR correspondence which has been performed in local quantum field theory \cite{Maldacena:1997re, Gubser:1998bc} in order to refine a quantum estimation of $\Lambda$. The UV/IR correspondence is applicable to several aspects of short vs. long distance physics and {\it ''deformed''} commutation relations. It is believed that the cosmological constant problem would reflect such a correspondence. 

To summarize, we utilize GUP; an extended version of HUP and an alternative quantum gravity approach, where a correction term encompassing the gravitational impacts \cite{Tawfik:2017syy,Tawfik:2016uhs,Dahab:2014tda, Ali:2013ma}. One features of the GUP exhibits the existence of a minimum length uncertainty \cite{Weinberg:1988cp,Banks:2000fe,Cohen:1998zx,ArkaniHamed:2000eg}.  There are so-far various approaches aiming at estimating the upper bound of the dimensionless GUP parameter $\beta_0$ by comparing such quantum gravity approaches to different phenomena, for example electroweak \cite{Das:2008kaa, Das:2009hs} and astronomical observations \cite{Scardigli:2014qka, Feng:2016tyt}. In the present work, we introduce a novel estimation of the $\beta_0$ parameter from binary neutron stars merger and gravitational wave event GW170817 \cite{TheLIGOScientific:2017qsa} as recently reported by LIGO and Advanced Virgo collaborations \cite{TheLIGOScientific:2017qsa}. Furthermore, we suggest an interpretation to the $\Lambda$ problem based on the UV/IR correspondence \cite{Chang:2001bm, Chang:2011jj,Miao:2013wua,Shababi:2017zrt, Vagenas:2019wzd}.

The present script is organized as follows. In section \ref{sec2}, we introduce the general notation of the quantum gravity approach, such as quadratic GUP. Section \ref{sec3} is devoted to a discussion on the energy-momentum dispersion relations and the speed of the graviton, at finite GUP. In the new representation of GUP, we show that the dimensionless GUP parameter, $\beta_o$, would be constrained by the gravitational wave event GW170817 \cite{TheLIGOScientific:2017qsa}. Section \ref{Liouville} constructs the Liouville theorem in a classical limit. Section \ref{sec5} presents a possible method to calculating the zero-point energy for the quantum harmonic oscillator and shows how this contributes to solving the cosmological constant problem with a quantum gravity approach; the GUP. The final conclusions are outlined in section \ref{sec6}.

%---------------------------------------------------------------
\section{The Approach: Generalized Uncertainty Principle (GUP) } 
\label{sec2} 
%---------------------------------------------------------------

Different approaches to the quantum gravity, such as GUP,  suggest the existence of a minimum length uncertainty of the order of Planck length \cite{Tawfik:2014zca, Tawfik:2015rva}. There are two main versions of the GUP, the quadratic GUP predicted from the string theory \cite{Amati:1988tn, Tawfik:2014zca, Tawfik:2015rva} and black hole physics  \cite{Maggiore:1993rv,Tawfik:2014zca, Tawfik:2015rva}, while the theory of doubly special relativity motivated linear GUP \cite{Magueijo:2002am, Tawfik:2014zca, Tawfik:2015rva}. The version of GUP proposed by \cite{Kempf:1994su} and supported by gedanken experiments dictates the length and momentum uncertainties
\bea
\Delta x\, \Delta p \geqslant \frac{\hbar}{2} \; \Big(1+\beta \left[  (\Delta p)^2 + \langle p\rangle^2 \right]\Big), \label{GUPEQ}
\eea
where $\beta$ is the GUP parameter, $\beta = \beta_0 (\ell_p/\hbar)^2 = \beta_0/ (M_p c)^2$, and $\beta_0$ is a dimensionless parameter to be estimated from cosmological observations, for instance. Eq. (\ref{GUPEQ}) exhibits the existence of a minimum length uncertainty, $\Delta x_{\mbox{min}} \approx \hbar \sqrt{\beta} =\ell_p \sqrt{\beta_0}$. $\ell_p=\sqrt{\hbar G/c^3}$ and $M_p=\sqrt{\hbar c/G}$ are Planck length and mass, respectively. We notice that $\Delta x$ is proportional to $\Delta p$, where large $\Delta p$, UV regime, becomes proportional to large $\Delta x$, IR regime. From the Jacobain identities, the GUP approach modifies the Heisenberg commutation relation,
 \bea
 \Big[ x_i,p_j \Big] = i\hbar \Big( \delta_{ij} (1+\beta p^2) + 2 \beta p_i p_j \Big), \label{GUPalgebra}
 \eea 
where $x_i=x_{0i}(1+\beta p^2)$, $p_j=p_{0j}$ and $p^2=\sum_{j=1}^{3}p^jp_j$. The operators $x_{0i}$ and $p_{0j}$ are similarly deduced from the corresponding noncommunitation relation $[x_{0i},p_{0j}]=\delta_{ij} i \hbar$.
 
%---------------------------------------------------------------
\section{Upper bound of GUP parameter from GW170817 }  
\label{sec3}
 %---------------------------------------------------------------
 
First, we need to derive the modified dispersion relation of the graviton due the GUP. The line metric with  Minkowski spacetime metric tensor $\eta_{\mu \nu}$ is given as 
 \bea
ds^2 =\eta_{\mu \nu} dx^\mu \,  dx^\nu = \eta_{00} c^2 dt^2 +  \eta_{ij} dx^i\,  dx^j.
\eea  
Accordingly, the modified four-momentum  reads
\bea
p_\mu p^\mu = \eta_{\mu \mu} p^\mu p^\mu  &=& \eta_{00} (p^0)^2 +\eta_{ij}  p^{0i}  p^{0j} (1+\beta p^2). \label{modifyMomentum}
\eea
In GR gravity, the energy of the graviton particle could be expressed as $\omega_g=-\eta_{00} c p^0$,
\bea
\omega_g^2 = m_g^2 c^4 + p^2 c^2 (1+ 2\beta p^2).   \label{MDRrel} 
\eea
The standard dispersion relation can be obtained when $\beta \rightarrow 0$. 

Second, we can now write down the modified speed of the graviton in presence of GUP. Toward this end, we investigate the upper bound of the dimensionless GUP parameter, $\beta_0$ taking into account conserving the violating Lorentz invariance \cite{Tawfik:2012hz} by estimating the speed of graviton from GW170817 \cite{TheLIGOScientific:2017qsa}. When assuming that the gravitational waves propagate as free waves through the medium, the speed of graviton can be determined from the group velocity of the accompanying wavefront   
\bea
v_g = \frac{ \partial \omega_g}{\partial p} = \frac{pc^2}{\omega_g} \Big( 1+ 4\beta p^2 \Big), \label{vgMDR1}
\eea
where $p$ is the canonical momentum of the graviton up to $\mathcal{O} (\beta)$. For   $p^2=(\omega_g/c)^2 - m^2 c^2 $,  Eq. (\ref{vgMDR1}) can be given as
\bea
v_g = c \;\left\{ \left[1-\left( \frac{ m_g c^2}{\omega_g}\right)^2 \right]^{1/2} +4\beta \frac{\omega_g^2}{c^2} \; \left[1-\left( \frac{  m_g c^2}{\omega_g}\right)^2 \right]^{3/2} \right\}. \label{vgraviton}
\eea
For $\beta \rightarrow 0$, the difference between the speed of light and that of graviton $\Big| \delta v\Big|$ can be obtained from GW170817 \cite{TheLIGOScientific:2017qsa}, where $m_g\lesssim 4.4 \times 10^{-22}~$eV$/c^2$ and $\omega = 8.5 \times 10^{-13}~$eV, respectively,
\bea
\Big| \delta v\Big| = \Big| c-v_g\Big| &=& \frac{1}{2}\left( \frac{ m_g c^2}{\omega_g}\right)^2 \lesssim 1.34 \times 10^{-19}\;c. \label{vDr}
\eea
For a massless graviton $ m_g \rightarrow 0$, this difference reads
\bea
\Big|\delta v_{\mbox{GUP}}\Big| &=&\Big| 4\beta \frac{\omega^2}{c}\Big|  \lesssim 1.95 \times 10 ^{-80} \beta_0 \;c, \label{vMDR}
\eea 
where $\beta = \beta_0/ (M_p c)^2$. %Instead of violating the  Lorentz invariance \cite{Tawfik:2012hz}. 

The upper bound on the dimensionless parameter, $\beta_0$, can be straightforwardly estimated from Eqs. (\ref{vDr}) and (\ref{vMDR}), 
\bea
\beta_0 \lesssim 8.89 \times 10^{60}. \label{MDRbeta1}
\eea  
This result is based on recent observations of mergers of spinning neutron stars and GW170817 \cite{TheLIGOScientific:2017qsa}. It intends to utilize this value of $\beta_0$ in solving the hierarchy problem of the vacuum energy; the cosmological constant problem.

%---------------------------------------------------------------
\section{Liouville theorem with minimal length uncertainty}  
\label{Liouville}
 %---------------------------------------------------------------
 
In this section, we shortly review the Liouville theorem in the classical limit \cite{Fityo:2008zz, Chang:2001bm, Wang:2010ct} to make sure that the size of each quantum state in the phase space volume is only depending on the momentum $p$. In other words, the number of quantum states in the phase space momentum is independent on time.

For noncommutative algebra of position and momentum operators, the Poisson bracket can be expressed as,
\bea
\left\{ F(x_1, \cdots x_D;\; p_1, \cdots p_D  ),  G(x_1, \cdots x_D;\; p_1, \cdots p_D  )\right\} =  \nn\\
\left(\frac{\partial F}{\partial x_i} \, \frac{\partial G}{\partial p_j} - \frac{\partial F}{\partial p_i} \frac{\partial G}{\partial x_j} \right) \left\{ x_i, p_j \right\} \nn \\ + \frac{\partial F}{\partial x_i} \frac{\partial G}{\partial x_j} \left\{ x_i, x_j \right\}. 
\eea   
During a time duration, $\delta t$, the Hamilton's equations of motion for position and momentum can be given as 
\bea
x_i^\prime = x_i + \delta x_i, \qquad \qquad  p_i^\prime = p_i + \delta p_i,
\eea
where $\dot{x}_i =  \delta x_i/  \delta t$ and $\dot{p}_i =  \delta p_i/  \delta t$ are  the time evolution of the length and momentum, respectively. Therefore, the equations of motion for length and momentum, respectively, reads
\bea 
 \delta x_i,    &=& \{ x_i, H\} \delta t = \{ x_i, p_j \}  \frac{\partial H}{\partial p_j} + \{ x_i, x_j\} \frac{H}{xj}, \\ 
 \delta p_i,    &=& \{ p_i, H\} \delta t = - \{ x_i, p_j \}  \frac{\partial H}{\partial x_j},
 \eea
where $H\equiv H(x,p;t)$ is the Hamiltonian.

The exchange in the phase space volume during $t$ time duration leads to determining the Jacobain of the transformation from $f(x_1, \cdots x_D;\; p_1, \cdots p_D)$ to $g(x_1^\prime, \cdots x_D^\prime;\; p_1^\prime, \cdots p_D^\prime)$, i.e.
\bea
d^Dx^\prime\; d^D p^\prime = \Big(d^Dx\; d^D p\Big)/\mathcal{J},
\eea
where $\mathcal{J}$ is the Jacobain of the transformation \cite{Fityo:2008zz, Chang:2001bm, Wang:2010ct},
\bea
\mathcal{J} &=& \Big\| \frac{\partial (x_1^\prime, \cdots x_D^\prime;\; p_1^\prime, \cdots p_D^\prime  )}{\partial (x_1, \cdots x_D;\; p_1, \cdots p_D  ) } \Big\|  =
 1 +  \left(\frac{\partial(\delta x_i)}{\partial x_i} + \frac{\partial(\delta p_i)}{\partial p_i}\right)  \nn \\ &=&   
1 + \left( \frac{\partial}{\partial x_i} \frac{\partial(\delta x_i)}{\partial t} + \frac{\partial}{\partial p_i} \frac{\partial(\delta p_i)}{\partial t} \right)  \delta t. 
\eea
The general notations of length and momentum brackets read
\bea
\big\{x_i, p_j\big\} &=& f_{ij} (x, p), \nn \\  \big\{x_i, x_j\big\} &=& g_{ij}(x,p),   \; \;  \mbox{and}    \nn \\     \big\{p_i,p_j\big\} &=& h_{ij}(p).
\eea
Thus, the Jacobain of the transformation \cite{Fityo:2008zz} can be rewritten as 
\bea 
\mathcal{J} = \prod_{i=1}^D f_{ii}(x,p)  =  1+ \sum_{i=1}^D (f_{ii} (x,p) - 1). \label{jacobian}
\eea
Accordingly, the invariant phase space in D-dimension is given as 
\bea
d^Dx^\prime\; d^D p^\prime = \Big(d^Dx\; d^D p\Big)/\Big(1+\beta p^2\Big)^D.
\eea
 Therefore, one can find the wight factor in 3D dimension for the quantum density of states as,
\bea
\frac{1}{(2\pi \hbar)^3} \frac{d^3 \vec{p}}{(1+\beta p^2)^3}.
\eea  

--------------------------------------------
\section{The cosmological constant problem} 
\label{sec5}
--------------------------------------------

The cosmological constant  $\Lambda$ can be related to the dark energy density $\Omega_\Lambda$ and the Hubble parameter $H_0$, $\Lambda=3\Omega_\Lambda\; H_0^2 $ \cite{Carroll:2000fy}. Also, the vacuum energy density can be expressed as,
\bea
\frac{c^2}{8 \pi G} \Lambda &=& \left(\frac{3 H_0^2 c^2}{8\pi G}\right) \Omega_\Lambda = \frac{3\hbar c}{8\pi \ell_p^2 \ell_0^2} \Omega_\Lambda,  \label{VacuEnergy}
\eea 
where $\ell_0$ is defined as the scale of visible light. From the recent updated data of PLANCK observations \cite{Aghanim:2018eyx}, the estimated values of $\Omega_\Lambda = 0.6889 \pm 0.0056$,  $H_0 = 67.66 \pm 0.42~$Km $\cdot$ s$^{-1}$ $\cdot$  Mpc$^{-1}$ \cite{Aghanim:2018eyx} and $\ell_0= c/H_0 = 1.368\times 10^{23}~$Km\cite{Aghanim:2018eyx}. Therefore, the vacuum energy density can be given in orders of magnitude of $10^{-47}~$GeV$^4/(\hbar^3c^3)$. 

From quantum theory, $\Lambda$ can be estimated from the sum over all momentum states of the vacuum fluctuation energies \cite{Carroll:2000fy}. The canonical energy of vacuum for each oscillator with a mass $m$ can be given as $(\hbar \omega_g)/2 = [p^2c^2+m_g^2c^4]^{1/2}/2$. For a massless particle, we find
\bea
\frac{1}{(2\pi \hbar)^3} \int d^3 p\; \frac{\hbar \omega_g }{2}) \simeq 9.60\times10^{74} \; {\mbox{GeV}}^4/ (\hbar^3 c^3).  \label{QFTlamda}
\eea
Accordingly, the cosmological constant problem can be now originated in the large gap between PLANCK obserrvations \cite{Aghanim:2018eyx} and Eq. (\ref{QFTlamda}). 

Concretely, when GUP is taken into account,  
\bea
\Lambda_{\mbox{GUP}} (m) &=& \frac{1}{(2\pi \hbar)^3} \int  d^3 p \; \rho(p^2)\;  \Big(\frac{1}{2}\hbar \omega_g \Big) \nn \\ &=& \frac{1}{2(2\pi \hbar)^3} \int   \frac{d^3 \vec{p}}{(1+\beta p^2)^3} \sqrt{p^2c^2+m_g^2c^4}. 
\eea
For a massless graviton particle, the vacuum energy density is given as
\bea
\Lambda_{\mbox{GUP}}(m\rightarrow0) &=& \ \frac{c}{4\pi^2 \hbar^3} \int \frac{p^3}{(1+\beta p^2)^3}\; dp =   \frac{c (M_p^2 c^2)^2}{16 \pi^2 \hbar^3 \beta_0^2} \nn \\ &=& 1.78 \times 10^{-48}~\mbox{GeV}^4/(\hbar^3 c^3). \label{GUPLamda}
\eea
Accordingly, the comparison between the observed value of $\Lambda$ \cite{Aghanim:2018eyx}, $\Lambda_{Obs} \approx 10^{-47}~\mbox{GeV}^4/(\hbar^3 c^3)$ and our calculations in  present of gravitational impacts, GUP, $\Lambda_{GUP} \approx 10^{-48}~\mbox{GeV}^4/(\hbar^3 c^3)$ seems very close. The potential match between the upper bound value of the GUP parameter as deduced from GW170817 \cite{TheLIGOScientific:2017qsa} and PLANCK 2018 \cite{Aghanim:2018eyx} tends to confirm the argument that the development of a quantum gravity theory is of considerable significance. One of such is an affordable interpretation of the vacuum catastrophe, i.e. the cosmological constant problem.

%-------------------------------------------------
 \section{Conclusions} 
 \label{sec6} 
%-------------------------------------------------

Taking into consideration the quadratic GUP, Eq. (\ref{GUPEQ}), the modified dispersion relations for GR were deduced, Eq. (\ref{MDRrel}). We have estimated the speed of graviton from the definition of the group velocity wavefront, Eq. (\ref{vgraviton}). Also, the difference between the speed of light and that of graviton, at vanishing $\beta$-parameter, i.e. vanishing GUP, Eq. (\ref{vDr}) and for massless graviton particle, Eq. (\ref{vMDR}), has been determined. The latter helped in calculating the upper bound on $\beta_0$-parameter, Eq. (\ref{MDRbeta1}) without violating the Lorentz invariance. We have shortly discussed on the Liouville theorem in the classical limit, where the number of quantum states in the phase space momentum remains constant with the time. 

We have introduced a novel estimation of the cosmological constant refining the disagreement shown between the quantum and observed values of $\Lambda$. Finally, we have obtained a good agreement between our estimated value and the observed one reported be the PLANCK collaboration \cite{Aghanim:2018eyx}.

The upper bound estimated for the dimensionless GUP parameter is $\beta_0 \approx 10^{60}$  which is based on the recent observations of mergers of spinning neutron stars. This result agrees well the ones reported in refs. \cite{Feng:2016tyt, Scardigli:2014qka} which are based to astronomical observations, as well. Moreover, We have applied the quantum gravity approach, GUP, and we made benefit from UV/IR features of the GUP in order to find a possible solution of the cosmological constant problem. We conclude that our theoretical result agrees within one order of magnitude with the observed one. 

We also conclude that the estimated upper bound of the GUP parameter, $\beta_0$, GW170817  \cite{TheLIGOScientific:2017qsa}, and its influence on the recently observed cosmological constant; PLANCK collaboration \cite{Aghanim:2018eyx}, might suggest a solution of one of the mysterious phenomena in physics, the cosmological constant problem.

\section*{Acknowledgments}
The authors are very grateful to thank the organizer committee of the $9^{\mbox{th}}$ International Workshop on Astronomy and Relativistic Astrophysics (IWARA $2020$ Video Conference) for the kind invitation.  
\bibliography{References}%

\end{document}